\documentclass[5p,twocolumn,10pt,times]{elsarticle}

\usepackage{hyperref}
\usepackage{appendix}
\usepackage{amsmath}
\usepackage{amsthm}
\usepackage{algorithmic}
\usepackage{algorithm}
\usepackage{soul}
\usepackage{xcolor}

\begin{document}

\baselineskip11pt

\begin{frontmatter}

\title{Indirect predicates for geometric constructions}


\author[first]{Marco Attene}
\ead{marco.attene@ge.imati.cnr.it}
\address[first]{CNR-IMATI, Via De Marni 6, 16149 Genova, Italy}

\begin{abstract}
Geometric predicates are a basic ingredient to implement a vast range of algorithms in computational geometry. Modern implementations employ floating point filtering techniques to combine efficiency and robustness, and state-of-the-art predicates are guaranteed to be always exact while being only slightly slower than corresponding (inexact) floating point implementations. Unfortunately, if the input to these predicates is an intermediate construction of an algorithm, its floating point representation may be affected by an approximation error, and correctness is no longer guaranteed.

This paper introduces the concept of \emph{indirect} geometric predicate: instead of taking the intermediate construction as an explicit input, an indirect predicate considers the primitive geometric elements which are combined to produce such a construction. This makes it possible to keep track of the floating point approximation, and thus to exploit efficient filters and expansion arithmetic to exactly resolve the predicate with minimal overhead with respect to a naive floating point implementation.
As a representative example, we show how to extend standard predicates to the case of points of intersection of linear elements (i.e. lines and planes) and show that, on classical problems, this approach outperforms state-of-the-art solutions based on lazy exact intermediate representations.
\end{abstract}


\end{frontmatter}


\section{Introduction}
The study and design of computational geometry algorithms is widely based on the so-called Real Arithmetic Model. According to that model, coordinates, lengths, angles, and many other entities, are all considered real numbers, and the algorithm correctness is based on this assumption. Practical implementations, however, are typically based on floating point (FP) arithmetic, and the aforementioned entities are represented by FP numbers. Floating point operations produce approximated results and, in some cases, the mismatch with the theoretical exact result might put the algorithm in an inconsistent state, cause infinite loops, and lead to crashes \cite{classroom} \cite{li2005}. An easy solution to this problem is to replace FP numbers with exact number types, but the consequent slowdown is rarely acceptable for practical applications.
Geometric algorithms have the peculiar property that their flow depends on the analysis of geometric entities and their configuration, and geometric predicates are the basic building blocks to perform such an analysis. Broadly speaking, a geometric predicate is a function that maps a set of geometric entities to an element of a (small) set of values. For example, the predicate \texttt{collinear} maps a triplet of points to a value in the set \{\texttt{true}, \texttt{false}\}, whereas the predicate \texttt{orient2d} maps a triplet of points to a value in the set \{\texttt{left\_turn}, \texttt{collinear}, \texttt{right\_turn}\}. To avoid inconsistencies, it is often essential that each predicate that determines the program flow produces exact results.

Arithmetic filtering techniques (see Sect. \ref{filtering_sota}) allow implementing predicates so that they are always exact, while being only slightly slower than a corresponding naive floating point implementation \cite{Devillers1998}. These techniques are appropriate when the predicate parameters are all input values and hence not computed as intermediate constructions by the algorithm. Conversely, if intermediate constructions are used as predicate parameters, an approximation during their computation may corrupt the result of the predicate, independently of how cleverly it has been implemented. Unfortunately, many useful algorithms require such intermediate constructions, and for these cases state-of-the-art robust solutions mostly rely on exact or lazy exact arithmetic kernels, which are still significantly slower and more memory intensive than floating point implementations.

In this paper, we show that most of these algorithms can be robustly implemented thanks to a novel concept of indirect geometric predicate. In particular, we show that the geometric construction itself can be filtered along with the predicate, thus leading to provably robust algorithms which are comparable with naive floating point implementations in terms of both performances and memory footprint.
The reference implementation and the scripts to reproduce the results in the paper are provided in the additional material and will be released as an open-source project.

\section{Background and prior art}
The majority of geometric predicates evaluate the sign of a (homogeneous) polynomial, typically a matrix determinant. As an example, the Cartesian 2D orientation of three points $a$, $b$, and $c$ corresponds to the sign of the determinant of the $2\times2$ matrix whose rows are the vectors $b-a$ and $c-a$. Instead of using exact arithmetic everywhere, which is too slow, modern implementations use arithmetic filtering. These implementations evaluate the polynomial sign using only FP arithmetic, but while keeping track of the approximation error. If this error is large enough to make the sign ambiguous (i.e. when the filter “fails”), they switch to exact arithmetic. The idea is that the failure rate is low enough to make the impact of exact arithmetic virtually negligible. Several approaches exist to bound the accumulated error. In general, calculating a precise bound is slower but minimizes the probability of filter failures. Conversely, conservatively approximating the bound is faster but less precise.

\subsection{Static, semi-static and dynamic filtering}
\label{filtering_sota}
Given an arithmetic expression involving a number of variables (e.g. the polynomial of a predicate), its evaluation using FP arithmetic is subject to a roundoff error that depends on the values assumed by the variables. This error can be bounded by a constant threshold that, in turn, can be computed analytically and initialized at compilation time. At runtime, if the magnitude of the evaluated expression is larger than this threshold, its sign is guaranteed to be correct. This approach is known as \emph{static filtering} \cite{fortune95}, and the overhead with respect to a plain FP implementation is due to a simple comparison. Without any assumption on the input, the threshold is typically too large and hence not really useful in practice. Nonetheless, if all the potential input values are known at the beginning of the program, the threshold can be computed once for all and then used to filter all the calls to the predicate.

Since the actual input to a predicate is typically made of points which are close to each other, considering only these points to derive a "local" threshold leads to tighter error bounds and, hence, to less filter failures. This approach is known as \emph{semi-static filtering}, and the overhead is due to a few additional operations to compute the threshold at each predicate call \cite{meyer08}.

Instead of computing an explicit threshold, interval arithmetic can be used to tightly bound the exact result. Basically, the result of any single operation is rounded by both excess and defect, thus leading to an interval containing the exact result. All the operations accumulate these approximations into growing intervals. If the interval representing the eventual result does not contain the zero, then its sign is defined without ambiguity. This approach is an example of \emph{dynamic filtering}, and evaluating the predicate in this way takes from three to eight times longer than plain FP calculation \cite{bronnimann98}.

It is often advantageous to use all these approaches in a cascaded multi-stage evaluation. Thus, the first attempt is based on plain FP calculation, and a single additional comparison allows implementing a static filter. If this filter fails, a few more comparisons (e.g. four for the \texttt{orient2d} predicate) allow implementing a semi-static filter. If this filter also fails, the polynomial is re-evaluated using interval arithmetic, which is the last attempt. As mentioned above, the first (static) filter makes sense only if a reasonably tight bound can be computed based on the input. In general, the process may safely start from the semi static filter while keeping the overhead reasonably small.

\subsection{Exact and lazy exact evaluation}
When all the filters fail, the predicate expression must be evaluated using exact arithmetic. State-of-the-art libraries allow, for example, to represent intermediate results using rational numbers, thus without any loss of precision \cite{Fortune93}. Also, polynomials can be exactly evaluated by using arbitrary precision floating point numbers \cite{Fousse2007} or so-called floating point expansions \cite{joldes2016}. This latter approach is particularly efficient as it fully exploits the floating point hardware. Shewchuck's geometric predicates \cite{shewchuk97} and Levy's Geogram \cite{Levy2016} exploit floating point expansions and are nearly as efficient as plain FP implementations on classical test beds. At the time of writing, Geogram predicates can be considered to represent the state of the art.

When intermediate constructions are used as input to a predicate, current robust solutions are based on lazy exact evaluation \cite{PION2011}. This is the approach adopted in \texttt{CGAL}'s Lazy Exact kernel: any intermediate construction is symbolically represented by the history of all the operations that combine input values to build the construction \cite{schirra14}. Besides keeping track of the operations, the construction is also approximately evaluated using interval arithmetic \cite{bronnimann98}. Since predicate expressions are also evaluated using interval arithmetic, their parameters can be intermediate constructions whose intervals propagate throughout the predicate's expression evaluation. If the resulting interval contains the zero, it means that the sign is ambiguous, and all the involved symbolic representations are tracked back and explicitly evaluated using exact arithmetic. The expectation is that in most of the cases interval arithmetic is sufficiently precise to reach an unambiguous conclusion, and hence exact evaluation is necessary only in a minority of the cases. Nonetheless, even in the ideal case where exact arithmetic is never called into play, interval arithmetic can be up to eight times slower than FP arithmetic \cite{bronnimann98}, and keeping track of the whole history has its own cost too.

Note that all the mechanisms described so far have been designed to compensate for numerical imprecision. Nonetheless, numerical imprecision is not the only source of ambiguity in geometric computation: many algorithms assume that their input is in so-called \emph{general position} and, if this is not the case, this precondition needs to be simulated \cite{edelsbrunner90}. This is rather orthogonal to compensating for numerical imprecision, and the two techniques should be combined when necessary.

\section{Geometric algorithm classification} \label{sec:algorithm_classes}
Herewith, we classify geometric algorithms based on the number of steps that require intermediate constructions. There are algorithms that have no such steps, that is, where all the predicates depend directly on input values. Then, there are algorithms having only one such step, where predicates depend either on input values or on intermediate constructions that, in their turn, depend directly on input values. And so on, up to arbitrarily complicated algorithms with cascaded sequences of predicate evaluations and intermediate constructions that mutually depend on each other. Our key observation is that the vast majority of state-of-the-art practical algorithms belong to the first two classes.

If $S$ is a set of 2D points whose coordinates are FP numbers, calculating their Delaunay triangulation amounts to repeatedly evaluate \texttt{orient2d} and \texttt{incircle} predicates whose input are points in $S$. Thus, this is a typical example of algorithm of the first class. Other examples include calculating convex hulls, Voronoi diagrams, surface interpolation, and many others.

If $S$ is a set of 2D segments that possibly intersect, calculating the constrained Delaunay triangulation of their arrangement is an example of algorithm of the second class. Indeed, the intersection of two segments is an intermediate construction used as input by the \texttt{orient2d} and \texttt{incircle} predicates. Other examples include the calculation of Minkowski sums \cite{Hachenberger2009}, mesh booleans \cite{Zhou2016}, exact mesh repairing \cite{attene2018}, and so on.

Current techniques allow combining robustness and efficiency only for algorithms of the first class thanks to arithmetic filtering. In this paper, we propose a novel approach to move forward and include algorithms of the second class too, thus covering a much wider subset of fundamental geometric algorithms.

\section{Indirect geometric predicates}
The core intuition behind our contribution is the following: if the code to calculate the intermediate construction is included in the predicate itself, arithmetic filtering may be applied to that code as well, and the predicate can be evaluated using FP arithmetic as long as the filter does not fail.

When intermediate constructions are used as parameters for a predicate, they are passed to the predicate in an unevaluated form: in this case, we say that these parameters are \emph{implicit} and the predicate is \emph{indirect}. 
An indirect predicate must accept at least one implicit parameter, but may accept standard \emph{explicit} parameters too.
In the latter case, the parameter has an explicit value and can be assumed to be exact (e.g. it is an input value or the result of an operation that does not introduce approximation errors). Conversely, in the former case the parameter is defined as an unevaluated function of explicit values (e.g. the average of input values). Since most geometric predicates have points as parameters, we extend this terminology to points and say that an explicit point is represented by its explicit coordinates, whereas an implicit point is represented by a geometric construction over explicit points (e.g. by the intersection of two lines, each defined by two explicit points).

The idea is to rewrite the expression of an implicit parameter $i$ as a fraction $\lambda/d$ where both the numerator $\lambda$ and the denominator $d$ are polynomials. Then, any standard predicate can be rewritten by substituting, in its own polynomial $\Lambda$, explicit parameters with polynomial fractions representing implicit parameters, thus reducing the whole predicate expression to another polynomial fraction $\Lambda'/D'$. Evaluating the sign of such a fraction amounts to evaluate the sign of both the numerator $\Lambda'$ and denominator $D'$, and both can be done robustly and efficiently through arithmetic filtering.

When an implicit parameter is a point, we derive a common denominator for all the fractions representing its Cartesian coordinates. Hence, an implicit point $i = (i_x, i_y, i_z)$ is rewritten as $i = (\lambda_{ix}/d_i, \lambda_{iy}/d_i, \lambda_{iz}/d_i)$ or, using homogeneous coordinates, as $(\lambda_{ix}, \lambda_{iy}, \lambda_{iz}, d_i)$.

Note that possible singular/degenerate configurations must be detected and appropriately handled. Typically, this amounts to check whether $d$ becomes zero for each of the implicit input values.

In the following subsections, we show how to apply this transformation to some non-trivial geometric constructions, and how to apply them to fundamental geometric predicates.

\subsection{Line-line intersections in 2D} \label{sec:lambda_LLI}
Let $\textbf{p}$ be an implicit 2D point defined as the intersection of two lines $l_a$ and $l_b$, each defined by two points. Specifically, $\textbf{e}_{a1}$ and $\textbf{e}_{a2}$ define $l_a$, whereas $\textbf{e}_{b1}$ and $\textbf{e}_{b2}$ define $l_b$ (Fig. \ref{fig:intersections}-left). Therefore:

\begin{align*}
\textbf{p} = (\frac{\lambda_x}{d} , \frac{\lambda_y}{d})
\end{align*}

where:

\begin{align*}
\lambda_x & = (e_{a1x} e_{a2y} - e_{a2x} e_{a1y} )(e_{b1x} - e_{b2x} ) + \\
			& -(e_{b1x} e_{b2y} - e_{b2x} e_{b1y} )(e_{a1x} -e_{a2x} )\\
\lambda_y & = (e_{a1x} e_{a2y} - e_{a2x} e_{a1y} )(e_{b1y} - e_{b2y} ) + \\
			& -(e_{b1x} e_{b2y} - e_{b2x} e_{b1y} )(e_{a1y} -e_{a2y} )\\
d & = (e_{a1x} -e_{a2x} )(e_{b1y} - e_{b2y} )-(e_{a1y} -e_{a2y} )(e_{b1x} - e_{b2x} )
\end{align*}

Note that $\textbf{p}$ is undefined if $d = 0$, which happens if $l_a$ and $l_b$ are parallel or if either of the two point pairs is degenerate.

\subsection{Line-plane intersections in 3D} \label{sec:lambda_LPI}
\begin{figure}
\includegraphics[width=0.8\linewidth]{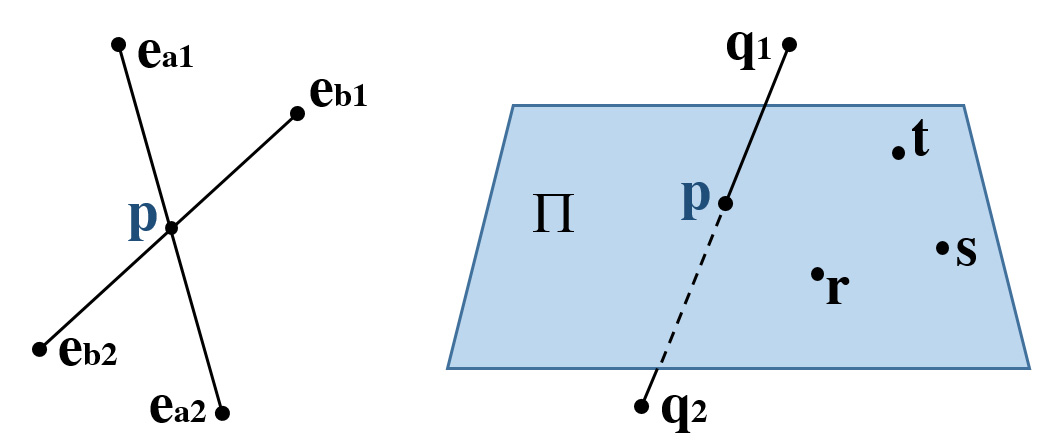}
\caption{Implicit point examples. Left - line-line intersection: four explicit 2D points define the implicit point \textbf{p}. Right - line-plane intersection: five explicit 3D points define the implicit point \textbf{p}.}
\label{fig:intersections}
\end{figure}

Let $\textbf{p}$ be an implicit 3D point defined as the intersection of a straight line $L$ and a plane $\Pi$, let $L$ be defined by two points $\textbf{q}_1$, $\textbf{q}_2$, and let $\Pi$ be defined by three points $\textbf{r}$, $\textbf{s}$, $\textbf{t}$ in the Euclidean space (Fig. \ref{fig:intersections}-right). Therefore:

\begin{align*}
\textbf{p} = (\frac{\lambda_x}{d} , \frac{\lambda_y}{d} , \frac{\lambda_z}{d})
\end{align*}

where

\begin{align*}
d =
\begin{vmatrix}
\textbf{q}_1 - \textbf{q}_2 \\
\textbf{s} - \textbf{r} \\
\textbf{t} - \textbf{r}
\end{vmatrix}
, n =
\begin{vmatrix}
\textbf{q}_1 - \textbf{r} \\
\textbf{s} - \textbf{r} \\
\textbf{t} - \textbf{r}
\end{vmatrix} \\
\lambda_x = d q_{1x} + n q_{2x} - n q_{1x} \\
\lambda_y = d q_{1y} + n q_{2y} - n q_{1y} \\
\lambda_z = d q_{1z} + n q_{2z} - n q_{1z}
\end{align*}

Note that $\textbf{p}$ is undefined if $d = 0$, which happens if $L$ and $\Pi$ are parallel, if  $\textbf{q}_1 = \textbf{q}_2$, or if $\textbf{r}$, $\textbf{s}$, $\textbf{t}$ are collinear or coincident.

\subsection{orient2d} \label{sec:orient2d}
Let $\textbf{p}_1$, $\textbf{p}_2$, $\textbf{p}_3$ be three points on the Euclidean plane. The polynomial expression of the classical \texttt{orient2d} predicate is $\Lambda = det \vert \textbf{p}_2 - \textbf{p}_1 ; \textbf{p}_3 - \textbf{p}_1 \vert = (p_{2x}-p_{1x}) (p_{3y}-p_{1y}) - (p_{2y}-p_{1y}) (p_{3x}-p_{1x})$. To start with, consider the simple case where just one of the points is an intermediate construction. Let it be $\textbf{p}_1$, and let us assume that we are able to rewrite its Cartesian coordinates as polynomial fractions (e.g. as described in Sect. \ref{sec:lambda_LLI}). Therefore:

\begin{align*}
p_{1x}=\frac{\lambda_x}{d} , p_{1y}=\frac{\lambda_y}{d}
\end{align*}

By substituting $\textbf{p}_1$ with $(\lambda_x/d, \lambda_y/d)$, we obtain $\Lambda = (p_{2x}-\lambda_x/d) (p_{3y}-\lambda_y/d) - (p_{2y}-\lambda_y/d) (p_{3x}-\lambda_x/d) = \Lambda'/D'$, where $\Lambda' = (d p_{2x}-\lambda_x) (d p_{3y}-\lambda_y) - (d p_{2y}-\lambda_y) (d p_{3x}-\lambda_x)$ and $D'=d^2$. Since the denominator is a square, it can be safely dropped and the sign is unambiguously defined by $\Lambda'$.

Now, let us consider the most generic case where all the three points are intermediate constructions. We obtain:

\begin{align*}
&p_{1x}=\frac{\lambda_{1x}}{d_1} , p_{1y}=\frac{\lambda_{1y}}{d_1}\\
&p_{2x}=\frac{\lambda_{2x}}{d_2} , p_{2y}=\frac{\lambda_{2y}}{d_2}\\
&p_{3x}=\frac{\lambda_{3x}}{d_3} , p_{3y}=\frac{\lambda_{3y}}{d_3}
\end{align*}

Where the $\lambda_i$ and the $d_i$ are computed as above. By substituting $\textbf{p}_1$, $\textbf{p}_2$, $\textbf{p}_3$ with their corresponding fractions, we obtain $\Lambda = (\lambda_{2x}/d_2-\lambda_{1x}/d_1) (\lambda_{3y}/d_3-\lambda_{1y}/d_1) - (\lambda_{2y}/d_2-\lambda_{1y}/d_1) (\lambda_{3x}/d_3-\lambda_{1x}/d_1) = \Lambda' / D'$, where

\begin{align*}
&\Lambda' = (d_1\lambda_{2x}- d_2\lambda_{1x}) (d_1\lambda_{3y} - d_3\lambda_{1y}) - (d_1\lambda_{2y}- d_2\lambda_{1y}) (d_1\lambda_{3x} - d_3\lambda_{1x}) \\
&D' = d_1^2 d_2 d_3.
\end{align*}

\subsection{incircle}
Let $\textbf{p}_1$, $\textbf{p}_2$, $\textbf{p}_3$, $\textbf{p}_4$ be four points on the Euclidean plane. The polynomial expression of the classical \texttt{incircle} predicate is the following:

\begin{align*}
\Lambda =
\begin{vmatrix}
p_{1x}-p_{4x} & p_{1y}-p_{4y} & (p_{1x}-p_{4x})^2+(p_{1y}-p_{4y})^2 \\ 
p_{2x}-p_{4x} & p_{2y}-p_{4y} & (p_{2x}-p_{4x})^2+(p_{2y}-p_{4y})^2 \\ 
p_{3x}-p_{4x} & p_{3y}-p_{4y} & (p_{3x}-p_{4x})^2+(p_{3y}-p_{4y})^2
\end{vmatrix}
\end{align*}

If each of the four points is implicitly defined, we can rewrite this expression as:

\begin{align*}
\Lambda =
\begin{vmatrix}
\frac{\lambda_{1x}}{d_1} - \frac{\lambda_{4x}}{d_4} & \frac{\lambda_{1y}}{d_1} - \frac{\lambda_{4y}}{d_4} & (\frac{\lambda_{1x}}{d_1} - \frac{\lambda_{4x}}{d_4})^2 + (\frac{\lambda_{1y}}{d_1} - \frac{\lambda_{4y}}{d_4})^2\\
\frac{\lambda_{2x}}{d_2} - \frac{\lambda_{4x}}{d_4} & \frac{\lambda_{2y}}{d_2} - \frac{\lambda_{4y}}{d_4} & (\frac{\lambda_{2x}}{d_2} - \frac{\lambda_{4x}}{d_4})^2 + (\frac{\lambda_{2y}}{d_2} - \frac{\lambda_{4y}}{d_4})^2\\
\frac{\lambda_{3x}}{d_3} - \frac{\lambda_{4x}}{d_4} & \frac{\lambda_{3y}}{d_3} - \frac{\lambda_{4y}}{d_4} & (\frac{\lambda_{3x}}{d_3} - \frac{\lambda_{4x}}{d_4})^2 + (\frac{\lambda_{3y}}{d_3} - \frac{\lambda_{4y}}{d_4})^2
\end{vmatrix}
\end{align*}

The common denominator to all the nine elements is $D' = (d_1 d_2 d_3 d_4 )^2$. When none of the $d$'s is zero, the sign of $\Lambda$ corresponds to the sign of the following $\Lambda'$:

\begin{align*}
&\Lambda' =
\begin{vmatrix}
m_{1,1} & m_{1,2} & m_{1,3} \\
m_{2,1} & m_{2,2} & m_{2,3} \\
m_{3,1} & m_{3,2} & m_{3,3}
\end{vmatrix} \\
m_{1,1} & = d_1 d_4^2 \lambda_{1x} - d_1^2 d_4 \lambda_{4x} \\
m_{1,2} & = d_1 d_4^2 \lambda_{1y} - d_1^2 d_4 \lambda_{4y} \\
m_{1,3} & = d_4^2 (\lambda_{1x}^2 + \lambda_{1y}^2) + d_1^2 (\lambda_{4x}^2 + \lambda_{4y}^2) - 2 d_1 d_4 (\lambda_{1x} \lambda_{4x} + \lambda_{1y} \lambda_{4y}) \\
m_{2,1} & = d_2 d_4^2 \lambda_{2x} - d_2^2 d_4 \lambda_{4x} \\
m_{2,2} & = d_2 d_4^2 \lambda_{2y} - d_2^2 d_4 \lambda_{4y} \\
m_{2,3} & = d_4^2 (\lambda_{2x}^2 + \lambda_{2y}^2) + d_2^2 (\lambda_{4x}^2 + \lambda_{4y}^2) - 2 d_2 d_4 (\lambda_{2x} \lambda_{4x} + \lambda_{2y} \lambda_{4y}) \\
m_{3,1} & = d_3 d_4^2 \lambda_{3x} - d_3^2 d_4 \lambda_{4x} \\
m_{3,2} & = d_3 d_4^2 \lambda_{3y} - d_3^2 d_4 \lambda_{4y} \\
m_{3,3} & = d_4^2 (\lambda_{3x}^2 + \lambda_{3y}^2) + d_3^2 (\lambda_{4x}^2 + \lambda_{4y}^2) - 2 d_3 d_4 (\lambda_{3x} \lambda_{4x} + \lambda_{3y} \lambda_{4y})
\end{align*}


\subsection{orient2d3d}
Let $\Pi$ be an oriented plane identified by three non-aligned points $\textbf{r}$, $\textbf{s}$ and $\textbf{t}$ in the Euclidean 3D space. Also, let $\textbf{p}_1$, $\textbf{p}_2$, $\textbf{p}_3$ be three points on $\Pi$. The \texttt{orient2d3d} predicate generalizes the standard \texttt{orient2d} predicate, and states whether $\textbf{p}_1$, $\textbf{p}_2$, $\textbf{p}_3$ form a right turn, a left turn, or a straight line on $\Pi$. It turns out that $\texttt{orient2d3d}(\textbf{p}_1, \textbf{p}_2, \textbf{p}_3, \textbf{r}, \textbf{s}, \textbf{t}) = \texttt{orient2d}(\Gamma(\textbf{p}_1), \Gamma(\textbf{p}_2), \Gamma(\textbf{p}_3))*\texttt{orient2d}(\Gamma(\textbf{r}), \Gamma(\textbf{s}), \Gamma(\textbf{t}))$, where $\Gamma(\textbf{p}_i)$ indicates the 2D projection of $\textbf{p}_i$ on one of the three axis-aligned planes such that $\texttt{orient2d}(\Gamma(\textbf{r}), \Gamma(\textbf{s}), \Gamma(\textbf{t})) \neq 0$. An effective way to select an appropriate projection plane is to consider a normal vector at $\Pi$ and simply drop the coordinate with largest magnitude. Without loss of generality, from now on we assume that we are dropping the Z coordinate or, equivalently, that we are projecting on the XY plane (i.e. $\Gamma(<p_x, p_y, p_z>) = <p_x, p_y>$).

Now, let us suppose that $\textbf{p}_1$ is implicitly defined (e.g. as described in Sect. \ref{sec:lambda_LPI}). Therefore:

\begin{align*}
p_{1x}=\frac{\lambda_x}{d} , p_{1y}=\frac{\lambda_y}{d} , p_{1z}=\frac{\lambda_z}{d}
\end{align*}

Since we are projecting on XY, we do not even need to calculate $\lambda_z$, and simply re-write
$\Lambda = (p_{2x}-p_{1x}) (p_{3y}-p_{1y}) - (p_{2y}-p_{1y}) (p_{3x}-p_{1x}) = (p_{2x}-\lambda_x/d) (p_{3y}-\lambda_y/d) - (p_{2y}-\lambda_y/d) (p_{3x}-\lambda_x/d) = \Lambda'/D'$, where $\Lambda' = (d p_{2x}-\lambda_x) (d p_{3y}-\lambda_y) - (d p_{2y}-\lambda_y) (d p_{3x}-\lambda_x)$ and $D'=d^2$. Hence, $\texttt{orient2d}(\Gamma(\textbf{p}_1), \Gamma(\textbf{p}_2), \Gamma(\textbf{p}_3)) = \texttt{sign}(\Lambda')$.

\subsection{orient3d}
Similar arguments hold for the \texttt{orient3d} predicate. When all its input points are explicit, the \texttt{orient3d} polynomial is:

\begin{align*}
\Lambda =
\begin{vmatrix}
p_{1x} - p_{4x} & p_{1y} - p_{4y} & p_{1z} - p_{4z} \\
p_{2x} - p_{4x} & p_{2y} - p_{4y} & p_{2z} - p_{4z} \\
p_{3x} - p_{4x} & p_{3y} - p_{4y} & p_{3z} - p_{4z}
\end{vmatrix}
\end{align*}

If we replace $\textbf{p}_1$ with an implicit point we obtain the following expression for an indirect \texttt{orient3d} predicate:

\begin{align*}
\Lambda'' =
\begin{vmatrix}
\lambda_x - d p_{4x} & \lambda_y - d p_{4y} & \lambda_z - d p_{4z} \\
d p_{2x} - d p_{4x} & d p_{2y} - d p_{4y} & d p_{2z} - d p_{4z} \\
d p_{3x} - d p_{4x} & d p_{3y} - d p_{4y} & d p_{3z} - d p_{4z}
\end{vmatrix}
\end{align*}

which, if $d \neq 0$, has the same sign as the simpler:

\begin{align*}
\Lambda' =
\begin{vmatrix}
\lambda_x - d p_{4x} & \lambda_y - d p_{4y} & \lambda_z - d p_{4z} \\
p_{2x} - p_{4x} & p_{2y} - p_{4y} & p_{2z} - p_{4z} \\
p_{3x} - p_{4x} & p_{3y} - p_{4y} & p_{3z} - p_{4z}
\end{vmatrix}
\end{align*}

where the corresponding $D'$ is $d$.

Formulas for the case where more than one point is implicit can be easily derived using a similar rewriting.

\section{Computational models and implementation}
The input to an indirect predicate is a set of points, each being either explicit or implicit. In the former case, the point is a standard array of Cartesian coordinates, each encoded as a double precision FP number; in the latter case, it is a collection of explicit points to be combined. Upon invocation, an indirect predicate determines which of its input parameters are implicit, and internally calculates $\lambda$ and $d$ for each of them.
In our implementation we consider three computation models: floating point arithmetic (with semi-static filter), interval arithmetic (with dynamic filter), expansion arithmetic (exact). The predicate is evaluated with the fastest model which guarantees exactness.

\subsection{Filtered evaluation in floating point}
We begin by calculating the $d$'s and, for each of them, use a semi-static filter that states whether its sign is unambiguous. If the filter succeeds for all the $d$'s, we compute the $\lambda$'s, evaluate $\Lambda'$, and use another semi-static filter to guarantee that its sign is correct. If the former filter fails for one of the $d$'s, it means that precision is insufficient to exclude that the implicit point is undefined (e.g. intersection of parallel lines). In this case, it makes no sense to proceed with this arithmetic model. As soon as a filter fails, we stop and switch to interval arithmetic (Sect. \ref{sec:intervaleval}). Upon success, the sign of $\Lambda'$ is possibly inverted depending on the sign of $D'$. We observe that $D'$ is just a product of the $d$'s, therefore we simply count how many of them have a negative sign (possible multiplicities must be considered). If they are an even number, we return the sign of $\Lambda'$, otherwise its inverse. Semi-static filters for the floating point model are calculated using a variation of \cite{meyer08} (see Appendix \ref{app:semi-static} for details). 

\subsection{Filtered evaluation using intervals} \label{sec:intervaleval}
When starting with this arithmetic model, we first set the CPU rounding mode to \texttt{towards\_infinity}. This is required to correctly and efficiently evaluate the subsequent interval arithmetic block. Then, we calculate the $d$'s and check that their sign is not ambiguous (i.e. their intervals do not contain the zero). If this dynamic filter succeeds for all the $d$'s, we compute the $\lambda$'s, evaluate $\Lambda'$, and check that the resulting sign is not ambiguous. As soon as one of the filters fail we stop and switch to exact arithmetic (Sect. \ref{sec:exacteeval}). If all the filters succeed, the sign of $\Lambda'$ is possibly inverted depending on the sign of the $d$'s. Upon termination (both with success or failure), we reset the rounding to the standard \texttt{to\_nearest} mode required by the other computational models.

\subsection{Exact evaluation using expansions} \label{sec:exacteeval}
When the predicate is computed using an exact arithmetic model there is a minor difference. Indeed, when computing the $d$'s it might happen that some of them are exactly zero. If so, the predicate cannot be evaluated because at least one of its implicit parameters is undefined. Conversely, if all the $d$'s are nonzero, we calculate $\Lambda'$ and check whether it is zero. If so, there is no need to check the $d$'s and the predicate returns zero. Otherwise, we possibly invert the sign as in the previous models.

Since $\Lambda'$ is a polynomial, we may exactly evaluate its sign using floating point expansions. Roughly speaking, an expansion is an array of FP numbers whose sum is the exact number being represented. Under certain assumptions, FP expansions can be efficiently summed, subtracted, and multiplied, all without error and while exploiting the floating point hardware \cite{joldes2016}.

\subsection{Caching}
If the same implicit point is used as input to many predicate calls, its $\lambda$'s and $d$ are recomputed many times. Therefore, upon the first occurrence it is worth storing their values for future reuse. For the sake of implementing filters, besides these values we must also store information to accumulate error bounds. For semi-static filters, this amounts to store an additional FP number (see Appendix \ref{app:semi-static}). For dynamic filters, this amounts to store the $\lambda$'s and $d$ using intervals.

Clearly, such a caching mechanism speeds up the computation at the expense of a higher memory usage. Our experiments (see Sect. \ref{sec:experiments}) show that caching $\lambda$'s and $d$'s for floating point and interval arithmetic models produces relevant speed ups while requiring a relatively low increase in memory footprint. Conversely, storing these values for the exact model is not advantageous: indeed, storing expansions for each point requires a significant amount of memory, and the time overhead due to the management of all this additional data is higher than what we save in the few switches which are required in typical cases (Table \ref{tab:results_cache}).
Note that pre-calculated intervals are also used in CGAL Lazy exact kernel \cite{cgalbook}, but the use of cached FP filters makes our predicate evaluaton significantly faster (Sect. \ref{sec:experiments}).

\section{Code generation}
\label{code_generation}
An indirect predicate must consider all the possible configurations of its input parameters. For example, the \texttt{orient2d} predicate has three points as parameters, and the type of each of these points can be either \emph{implicit} or \emph{explicit}. The most generic implementation may assume that all the points are implicit and, if any of them is explicit, its $\lambda$'s can be simply replaced with the Cartesian coordinates and its $d$ with 1. Though this approach is correct, such a naive implementation would lead to unnecessarily loose filters which, in turn, would require unnecessary switches to more precise models. Consequently, this implementation would be unnecessarily slow.

A much more efficient approach is based on the implementation of a specific instance of the predicate for each of the possible input configurations. If the three input points of our indirect \texttt{orient2d} predicate are all explicit, the predicate reduces to the standard \texttt{orient2d}, which is very efficient and accumulates small errors. If only the first point is implicit, we may use a dedicated implementation based on the $\Lambda$ and $D$ described earlier in Sect. \ref{sec:orient2d}. This is slightly less efficient than the previous version and produces a slightly higher error. And so on. Hence, an indirect \texttt{orient2d} predicate might be efficiently implemented using eight instances (EEE, EEI, EIE, EII, IEE, IEI, IIE, III, where the three characters indicate the type of each input point). Actually, many of these instances can be reduced to each other by transposing their input parameters and possibly inverting the resulting sign (e.g. \texttt{orient2d}(a, b, c) = \texttt{orient2d}(b, c, a) = -\texttt{orient2d}(a, c, b)). It is easy to verify that four versions (EEE, IEE, IIE, III) of \texttt{orient2d} are sufficient to cover all the cases.

Implementing even a single instance with proper semi-static and dynamic filtering, and with expansion-based evaluation, is a tedious and error-prone process. Fortunately, this process can be mostly automated \cite{Nanevski2003}. If the predicate is naively implemented as a C function using FP arithmetic, software tools exist \cite{meyer08} to parse such a function and produce another C function which includes a semi-static filter. These tools analyze the polynomial and calculate an upper bound for the roundoff error. In the produced code, the magnitude of the resulting determinant is compared with such a bound and, if smaller, the predicate returns a particular value indicating that the filter failed (e.g. \texttt{FPG\_UNCERTAIN}). Meyer and Pion's tool has been used and extended by Levy in the \emph{Predicate Construction Kit} \cite{Levy2016}, where both the semi-static filtered and the exact versions of the predicate are produced.

We have further extended upon this approach, and have implemented a software that parses a text-based formula representing the polynomial. Starting from such a formula, our tool produces three C++ functions, each implementing the formula using one of the arithmetic models described earlier. A further fourth function is generated that calls the other three in order and terminates as soon as one of them returns a guaranteed result.

\section{Experiments} \label{sec:experiments}
For the sake of experimentation, we have implemented a C++ library with an API defining a \texttt{genericPoint} class. A \texttt{genericPoint} can be specialized to be either 2D or 3D, implicit or explicit. The library also provides functions implementing both standard and indirect predicates. Each such function takes a number of \texttt{genericPoint} objects as input, internally determines their type, and properly handles all the cases as described in Sect. \ref{code_generation}. The library has been used as a basis to implement a simple Delaunay triangulation algorithm that can deal with \texttt{genericPoint} objects. Both the library and the triangulation algorithm have been compiled and tested on an MS Windows 10 OS using Visual C++ 2019. The objectives are measuring the efficiency of our predicates, the overhead of indirect predicates when compared to their direct versions, the impact of all the possible caching levels, and the improvement with respect to the state of the art. Tests were run on a standard PC equipped with an Intel Core i7-4770 with 16 Gb RAM. Based on this setting, the following experiments were run. Results are summarized in Tables \ref{tab:results11_12} and \ref{tab:results_other}.

\begin{table}
\centering
\begin{tabular}{ | l | c | c |}
\hline
\textbf{N. Points} 		&\textbf{Exp. 1.1}	   &\textbf{Exp. 2.1}\\
\hline \hline
1000		&0.000691 (2.74)	&0.000862 (2.77)\\ \hline
10000		&0.008698 (6.29)	&0.008864 (6.49)\\ \hline
100000		&0.117321 (27.99)	&0.12001 (25.56)\\ \hline
1000000	&1.79624 (268.53)	&1.82953 (284.55)\\ \hline
\end{tabular}
\caption{Results of our experiments 1.1 and 2.1. Elapsed time in seconds and, in parenthesis, peak memory usage in Mbytes.}
\label{tab:results11_12}
\end{table}

\begin{table*}
\centering
\begin{tabular}{ | l | c | c | c | c |}
\hline
\textbf{\% implicit} 		&\textbf{Exp. 1.2}	   &\textbf{Exp. 1.3} 		&\textbf{Exp. 2.2}	   &\textbf{Exp. 2.3}\\
\hline \hline
0	&1.79624 (268.53)	&2.63308 (268.36)	&1.82953 (284.55)	&2.64815 (284.37)\\ \hline
10	&2.23134 (292.84)	&5.0791 (292.99)	&2.56663 (321.96)	&7.14617 (322.95)\\ \hline
25	&2.82864 (328.84)	&8.45898 (329.3)	&3.41211 (377.43)	&16.6439 (379.75)\\ \hline
50	&3.62846 (389.14)	&14.0788 (390.61)	&4.24588 (469.68)	&38.3586 (472.40)\\ \hline
100	&3.81999 (509.93)	&27.1925 (511.29)	&4.29742 (654.52)	&75.0639 (657.06)\\ \hline
\end{tabular}
\caption{Results of our experiments 1.2, 1.3, 2.2 and 2.3. Elapsed time in seconds and, in parenthesis, peak memory usage in Mbytes.}
\label{tab:results_other}
\end{table*}

\subsection{Experiment 1.1 - Random explicit 2D points}
For this experiment, we created and triangulated a set of \texttt{genericPoint} objects, each specialized to represent an explicit 2D point. Their Cartesian coordinates are randomly sampled within the unit square $[0,1] \times [0,1]$. Since all the points are explicit, the \texttt{orient2d} and \texttt{incircle} predicates used by the algorithm always switch to the standard \emph{direct} version of the predicates. A parameter allows controlling the cardinality of the point set in the range [1, 1000000]. The objective of this test is to define a reference for the other tests, in order to evaluate the overhead due to the use of indirect predicates when compared with their classical direct version.

\subsection{Experiment 1.2 - Random mixed 2D points}
Herewith, our set of \texttt{genericPoint} objects still contains 2D points, but their type is mixed. A subset of them are explicit, whereas the others are implicitly defined as the intersection of line pairs. Explicit points are randomly scattered within the unit square as in the previous experiment. Also, line pairs are created so that their intersections are randomly scattered within the unit square. A parameter allows controlling the percentage of implicit points over a total of 1 million points constituting the entire set. In this case, both \texttt{orient2d} and \texttt{incircle} may either switch to the direct version or select the proper indirect predicate instance, depending on the type of the \texttt{genericPoint}s.

\subsection{Experiment 1.3 - Regular mixed 2D points}
This dataset of 1000000 \texttt{genericPoint} objects contains mixed 2D points as in Experiment 1.2 but, instead of being randomly scattered, they form a regular grid within the unit square. A subset of them are explicit, whereas the others are implicitly defined as the intersection of line pairs. Even in this case, a parameter allows controlling the percentage of implicit points over the total. This dataset stresses the predicates that are largely required to switch to exact arithmetic.

\subsection{Experiment 2.1 - Random explicit 3D points}
For this experiment, we created a set of \texttt{genericPoint} objects, each specialized to represent an explicit 3D point whose Cartesian coordinates are randomly sampled within the unit cube $[0,1] \times [0,1] \times [0,1]$. The Delaunay triangulation is calculated on the projection of these points on the XY plane. As for Experiment 1.1, the objective is to define a reference for the following Experiments 2.2 and 2.3.

\subsection{Experiment 2.2 - Random mixed 3D points}
Here, our set of \texttt{genericPoint} objects contains 3D points, both explicit and implicit, and the Delaunay triangulation is calculated on their projection on the XY plane. As in the previous experiment, the first half of the set is made of explicit points, whereas the other half are implicitly defined as the intersection of lines and planes. Explicit points are randomly scattered within the unit cube $[0,1] \times [0,1] \times [0,1]$. Line-plane pairs are created so that their intersections belong to the unit cube. A parameter allows controlling the percentage of implicit points over a total of 1 million points constituting the entire set.

\subsection{Experiment 2.3 - Regular mixed 3D points}
This dataset of 1000000 \texttt{genericPoint} objects contains mixed 3D points as in Experiment 2.2 but, instead of being randomly scattered, their projections on the XY plane form a regular grid within the unit square. Their Z coordinate is still randomnly selected in the interval [0,1]. A subset of the points are explicit, whereas the others are implicitly defined as the intersection of properly configured line-plane pairs. Even in this case, a parameter allows controlling the percentage of implicit points over the total. This dataset stresses the predicates that are largely required to switch to exact arithmetic.

\begin{table*}
\centering
\begin{tabular}{ | l | c | c | c | c |}
\hline
\textbf{\% implicit} 		&\textbf{no cache}	   &\textbf{FP} 		&\textbf{Interval}	   &\textbf{Exact}\\
\hline \hline
0	&1.79552 (268.51)	&1.78396 (268.53)	&1.79624 (268.53)	&1.82953 (268.53)\\ \hline
10	&2.49302 (283.07)	&2.4852 (286.20)	&2.23134 (292.84)	&2.27749 (358.66)\\ \hline
25	&3.5918 (304.77)	&3.39718 (312.93)	&2.82864 (328.84)	&2.86765 (494.35)\\ \hline
50	&5.25428 (341.23)	&5.00143 (357.23)	&3.62846 (389.14)	&3.79824 614.96)\\ \hline
100	&7.16629 (413.65)	&6.9816 (445.84)	&3.81999 (509.93)	&3.9589 (1167.84)\\ \hline
\end{tabular}
\caption{Results of our experiment 1.2 using different caching approaches. $\lambda$'s and $d$ were cached only for the FP model (column \textbf{FP}), for both the FP and interval arithmetic models (column \textbf{Interval}), or for all the available models (column \textbf{Exact}). Elapsed time in seconds and, in parenthesis, peak memory usage in Mbytes.}
\label{tab:results_cache}
\end{table*}

\subsection{Comparison with state of the art}
\texttt{CGAL} \cite{cgalbook} is probably one of the most used libraries when robustness and efficiency must be combined, in particular when important fundamental algorithms come into play (e.g. Delaunay triangulations). That is why we compare our results with \texttt{CGAL}. In our comparison, we always strive to exploit \texttt{CGAL} in its most efficient configuration while providing a guarantee of correctness. Namely, we want to guarantee that the resulting Delaunay triangulation has the correct topology.

We re-run all our tests using \texttt{CGAL} Delaunay triangulation algorithm. For experiments 1.1 and 2.1, where all the points are explicit, we configure \texttt{CGAL} to use the \texttt{Exact\_Predicates\_Inexact\_Constructions} kernel, whereas for the others we need exact constructions to represent the intersections. Implicit points are generated using \texttt{CGAL::intersection()}. For experiments that require projection on the XY plane, \texttt{CGAL::Projection\_traits\_xy\_3} was used.
We assume that \texttt{CGAL} implements efficient state-of-the-art algorithms but, since using two different implementations may introduce a bias independently of our contribution, we have also re-implemented our own triangulation algorithm using \texttt{CGAL}'s exact number type \texttt{CGAL::Lazy\_exact\_nt<CGAL::Gmpq>} and its predicates \texttt{CGAL::orientation()} and \texttt{CGAL::side\_of\_oriented\_circle()}. However, after having verified that this re-implementation is slower than \texttt{CGAL}'s native algorithm, we have abandoned it. Tables \ref{tab:cgal_1} and \ref{tab:cgal_2} report the results of our experiments using \texttt{CGAL}.

\begin{table}
\centering
\begin{tabular}{ | l | c | c |}
\hline
\textbf{N. Points} 		&\textbf{Exp. 1.1}	   &\textbf{Exp. 2.1}\\
\hline \hline
1000		&0.000253 (3.52)	&0.000273 (3.29)\\ \hline
10000		&0.015636 (4.68)	&0.015622 (4.72)\\ \hline
100000		&0.098743 (16.64)	&0.109367 (19.62)\\ \hline
1000000	&1.168691 (135.81)	&1.15617 (167.65)\\ \hline
\end{tabular}
\caption{Results of our experiments 1.1 and 2.1 using CGAL with exact predicates and inexact constructions kernel. Elapsed time in seconds and, in parenthesis, peak memory usage in Mbytes.}
\label{tab:cgal_1}
\end{table}

\begin{table*}
\centering
\begin{tabular}{ | l | c | c | c | c |}
\hline
\textbf{\% implicit} 		&\textbf{Exp. 1.2}	   &\textbf{Exp. 1.3} 		&\textbf{Exp. 2.2}	   &\textbf{Exp. 2.3}\\
\hline \hline
0	&6.92141 (168.86)	&26.4514 (362.26)	&33.2009 (193.40)	&68.7454 (483.43)\\ \hline
10	&8.09864 (252.48)	&27.9512 (492.32)	&35.0445 (338.69)	&76.5043 (716.04)\\ \hline
25	&9.76497 (377.75)	&32.0291 (698.53)	&38.7077 (553.73)	&88.9425 (1060.50)\\ \hline
50	&11.9211 (586.53)	&40.966 (1040.29)	&42.2784 (908.59)	&111.112 (1634.54)\\ \hline
100	&15.9442 (1006.73)	&60.9706 (1682.76)	&50.1841 (1605.84)	&crashed\\ \hline
\end{tabular}
\caption{Results of our experiments 1.2, 1.3, 2.2 and 2.3 using CGAL with exact predicates and exact constructions kernel. Elapsed time in seconds and, in parenthesis, peak memory usage in Mbytes.}
\label{tab:cgal_2}
\end{table*}

\subsection{Results discussion}
In all the experiments, the total number of predicate calls depends on the input size. For Experiment 1.1, these numbers are as follows: 1000 points $\Rightarrow$ 4395 calls to \texttt{orient2d} and 7502 calls to \texttt{incircle}; 10K points $\Rightarrow$ 45K and 94K calls; 100K points $\Rightarrow$ 469K and 1.14M calls; 1M points $\Rightarrow$ 4.8M and 13.3M calls. These numbers do not vary significantly across different experiments with same input size.

Tables \ref{tab:results11_12} and \ref{tab:results_other} show that the impact of having implicit points in the input is rather limited. The maximum increase in both time and memory usage is around 2x in 2D and 2.5x in 3D, which happens when all the points are implicit. Most importantly, the resource demand gradually increases as the relative amount of implicit points in the set grows. Hence, our approach is particularly convenient in cases where implicit points are sparse.
Experiments 1.3 and 2.3 show that our predicates behave reasonably well even in extremely degenerate cases where exact arithmetic is used extensively. In particular, we observe that the peak memory usage in these cases is virtually equivalent to that of corresponding non-degenerate input sets. The impact on the elapsed time is of course more significant but, even in this case, we observe that it gracefully grows as the percentage of implicit points increases.
Table \ref{tab:results_cache} shows the impact of caching the $\lambda$'s and $d$'s: when these values are stored for the FP model only the overall gain in speed is modest, but when the Interval model is included the algorithm becomes nearly twice as fast with only 23\% more memory requirements.

\begin{figure*}
\includegraphics[width=\linewidth]{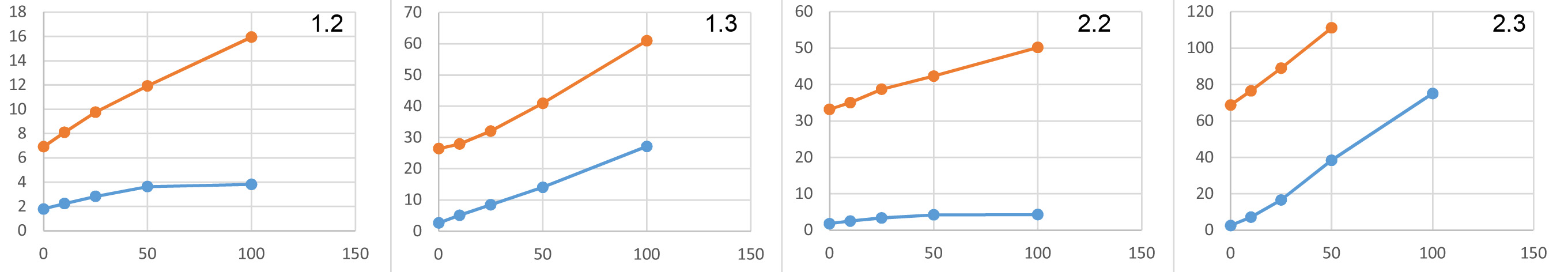}
\caption{Elapsed time for experiments 1.2, 1.3, 2.2 and 2.3 using indirect predicates (blue) and CGAL (orange).}
\label{fig:plots-time}
\end{figure*}

Our results also demonstrate that indirect predicates outperform state of the art methods with same guarantees. Even if \texttt{CGAL} is faster when inexact constructions are allowed (Experiments 1.1 and 2.1, see Tables \ref{tab:results11_12} and \ref{tab:cgal_1}), its lazy exact kernel becomes necessary to represent implicit points and, in this case, our implementation is from 3.86 to 11.7 times faster than \texttt{CGAL}, while requiring up to 3.5 times less memory (see Tables \ref{tab:results_other} and \ref{tab:cgal_2} and Fig. \ref{fig:plots-time}).

\section{Conclusions}
We have shown that, even if intermediate geometric constructions are necessary, the efficiency of floating point hardware can be exploited to quickly evaluate geometric predicates with guarantees of correctness. This partly solves a problem introduced by Meyer and Pion \cite{meyer08} more than ten years ago. The problem is only \emph{partly} solved because our approach requires that intermediate constructions can be represented as polynomial fractions. Though this is sufficient for most of the algorithms of the second type introduced in Sect. \ref{sec:algorithm_classes}, there are still a few useful algorithms that do not satisfy this requirement. One noticeable example is the constrained Delaunay triangulation in 3D, where state of the art algorithms use irrational numbers to generate Steiner points \cite{hangsi15}.
Furthermore, more \emph{numerically complex} techniques exist which involve cascaded predicates and constructions, and hence do not belong to our second class of algorithms (e.g. \cite{cazals08}). Unfortunately, indirect predicates do not help in these cases, and attempting to generalize their paradigm is a challenging research direction: indeed, the combination of cascaded constructions would easily lead to rather complicated expressions whose filters become too large to be useful in practice.
Nonetheless, indirect predicates enable much faster implementations for a wide range of useful algorithms of the second class.

We also observe that indirect predicates allow implementing algorithms with guarantees of robustness and combinatorial correctness. Nonetheless, if their results are the input to other algorithms which expect floating point coordinates, implicit points must be evaluated and their coordinates approximated. Such an approximation may lead to invalid configurations (e.g. with self intersections) even if the combinatorics is exact. This problem is known as \emph{snap rounding} and, though effective algorithms exist for the 2D case, no practical guaranteed implementation exists yet for the 3D case. This problem is worth a further investigation \cite{lazard18}.


\section*{Acknowledgements}
This work is partly supported by the EU ERC Advanced Grant CHANGE, agreement No 694515. Thanks are due to the colleagues at CNR-IMATI for helpful discussions.

\section*{References}
\bibliography{mybibfile}

\begin{appendices}

\section{Semi-static filter calculation}
\label{app:semi-static}
A semi-static filter needs to (1) calculate a threshold value as a function of input parameters $v_i$ and (2) verify that $\vert \Lambda' \vert$ is larger than this value. In our code, the threshold $\epsilon(v_1, ..., v_n)$ is calculated as a product $\delta(1)B(v_1, ..., v_n)$, where $\delta(1)$ is a constant value which depends only on the predicate's expression and is computed at compile time as described in \cite{meyer08}. Conversely, $B(v_1, ..., v_n)$ depends on the input values $v_i$ and must be calculated at each predicate call. This \emph{dynamic} part of the filter is the product of $k$ factors $b_1 b_2 \ldots b_k$, where $b_j$ is the absolute value of either one of the input variables $v_i$ or a difference $v_a - v_b$ between two of them. Note that factors need not be different from each other (e.g. $b_i$ and $b_j$ may be the same for $i \neq j$). Differently from \cite{meyer08}, we slightly overestimate $B(v_1, ..., v_n)$ for the sake of efficient caching. Specifically, we replace the product $b_1 b_2 \ldots b_k$ with the power $\beta^k$, where $\beta = argmax\{b_1, \ldots, b_k\}$. The reason for this is the following.

\begin{figure}
\includegraphics[width=\linewidth]{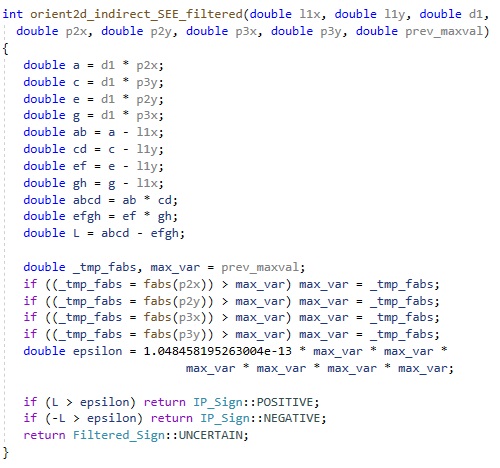}
\caption{Code produced by our tool for an indirect version of the \texttt{orient2d} predicate. The first of the three points is implicitly defined by the intersection of two lines that, in turn, are defined by explicit points $w_1, \ldots, w_4$. The values for $\lambda_x$, $\lambda_y$ and $d$ (\texttt{l1x}, \texttt{l1y}, \texttt{d1}) are calculated externally and passed to this function along with the maximum $\beta$ deriving from the $w$'s (\texttt{prev\_maxval}). The constant value of $\delta(1)$ ( = \texttt{1.048458195263004e-13}) accumulates the error bounds of this function and those of the external function that calculates the $\lambda$'s and $d$.}
\label{fig:filtercode}
\end{figure}

Let $P(v_1, \ldots, v_k, i_1, \ldots, i_n)$ be a predicate where the $v_i$'s are explicit parameters whereas the $i_j$'s are implicit. Any of the implicit parameters $i$ is defined as a composition of explicit values $w_1 \ldots w_m$. Upon evaluation using the FP model, we first use the $w$'s to calculate the $\lambda$'s and $d$'s, and then use these $\lambda$'s and $d$'s together with the explicit parameters $v$'s to evaluate the whole predicate.
The semi-static filter for this composed calculation must take all the variables into account. Thus, the dynamic component of our filter is $\beta^k$, where $\beta$ is the maximum of the $b_i$'s deriving from explicit variables $v$ and the $b_j$'s deriving from the $w$'s. The maximum over the $b_j$'s is computed only once for each implicit point and cached for later use. Hence, replacing the product with the power allows using a single variable to cache the bounds for any $\lambda$ and $d$. When compared with the caching benefits, the performance loss due to our overestimation is virtually negligible. Fig. \ref{fig:filtercode} shows an example of the code produced by our tool.

\end{appendices}

\end{document}